\documentclass[11pt]{article}
\usepackage{epsfig}
\usepackage{jheppub}
\usepackage{atlasphysics}
\pdfoutput=1 
\begin{document}

\begin{flushright}
25 October 2014
\end{flushright}

\vskip 1cm

\begin{center}
{\bf{\LARGE STUDY OF SINGLE TOP PRODUCTION }}
\end{center}
\begin{center}
{\bf{\LARGE AT HIGH ENERGY ELECTRON }}
\end{center}
\begin{center}
{\bf{\LARGE POSITRON COLLIDERS}}
\end{center}

\begin{center}
J. Fuster, I.Garc\'\i a, M.Perell\'o, E.Ros, M.Vos \\ 
\it{IFIC (Univ. Valencia - CSIC)}
\end{center}

\vskip 1cm

\begin{abstract}

\noindent

The effect of single top production on the study of top quark
pair production in future high energy electron--positron colliders 
is evaluated. 
The rate of the single top quark production process is sizeable throughout 
a large range of center-of-mass energies and cannot easily be distinguished
from the dominant pair production process. We discuss the impact
on the top quark mass extraction from a scan through the pair production
threshold and the determination of top quark form factors in the continuum.
These results advocate for the exploration of the inclusive
$e^+e^-\rightarrow W^+bW^-\bar{b}$ process, that includes both 
top quark pair and single top quark production.


\end{abstract}

\vskip 1cm

\section{Introduction}

A high-luminosity, high-energy, linear $e^+ e^-$ collider yields excellent 
opportunities for precision tests of the Standard Model of particle physics. 
The combination of precisely calculable electroweak production and 
strict control of the initial
state with the relatively benign experimental environment and 
state-of-the-art detector systems allow for a characterization
of Standard Model and new physics processes with a precision 
that goes well beyond what can be achieved at hadron colliders.

Two projects of linear electron-positron colliders are being considered: 
the International Linear Collider (ILC~\cite{Behnke:2013xla,BrauJames:2007aa}) and the Compact Linear Collider (CLIC~\cite{cliccdr1}). 
The physics case for a linear $e^+e^-$ machine has been made in great detail in 
References~\cite{Baer:2013cma,Brau:2012hv,Djouadi:2007ik,snowmass1,snowmass2,
snowmass3,snowmass4,tesla}. The specific case of a multi-\tev{} $ e^+ e^- $ 
collider is discussed in References~\cite{cliccdr2,cliccdr3,clicphysics}.
In both cases, the center-of-mass energy will exceed
$\sqrt{s}$= 350 \gev{}, the threshold for top quark pair production. 
Unlike other
quarks, the top quark has never been produced in  $e^+e^-$ machines, and 
therefore a precise measurement of electroweak top quark
pair production is missing. The study of  top quark properties is therefore
one of the most exciting prospects for a future linear collider~\cite{Asner:2013hla}. Detailed full-simulation studies have been made of the prospects 
for a precise top quark mass measurement~\cite{Seidel:2013sqa,Horiguchi:2013wra} and characterization of the
$t\bar{t}Z$ and $t\bar{t}\gamma$ vertices~\cite{Amjad:2013tlv}.

\begin{figure}[htb]
\begin{center}
  \includegraphics[width=0.3\linewidth]{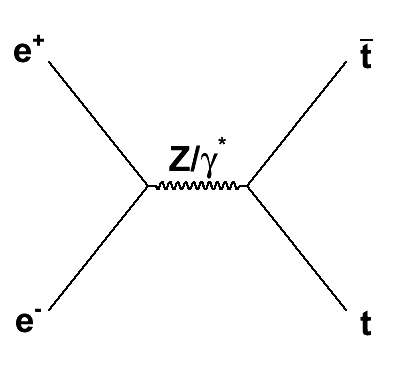} 
  \includegraphics[width=0.3\linewidth]{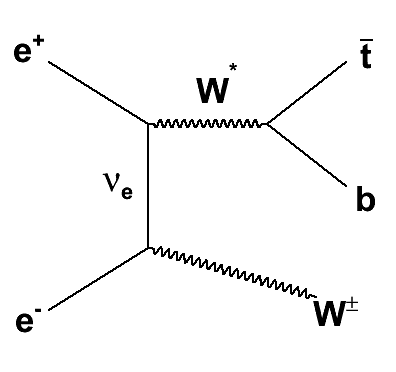}  
  \includegraphics[width=0.3\linewidth]{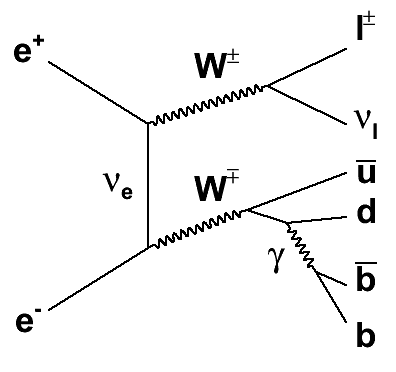}
  \caption{Feynman diagrams for top quark pair production at a linear collider 
   ($e^+e^⁻ \rightarrow Z/\gamma^* \rightarrow t\bar{t}$, left panel), single top 
   production ($e^+e^- \rightarrow W^{-} t\bar{b}, W^+ \bar{t} b$, central panel), and 
   triple gauge boson production ($e^+e^- \rightarrow W^+W^-Z$, right panel).           
    \label{fig:diagrams}}
\end{center} 
\end{figure}

Single top production, through the 
$e^+e^- \rightarrow W^{-} t\bar{b}, W^+ \bar{t} b$ process depicted in the 
central panel of Figure~\ref{fig:diagrams}, is abundant at $e^+ e^-$ 
colliders that operate at $\sqrt{s}>$300~\gev. 
Note that for the $t \rightarrow bW $ decay, this process gives rise to 
the same $W^+bW^-\bar{b}$ final state as top pair production. 
Using MADGRAPH we find ten leading-order diagrams of this type, against 
two for top quark production. 
A third group of processes gives rise to the same final state: 
$WWZ$, $WWh$ and $WW\gamma$ production, with $Z/\gamma/h \rightarrow b\bar{b}$. At tree-level there are approximately 50 diagrams that produce two $W$-bosons
and two $b$-quarks that do not involve top quarks.
 
Ultimately, all three processes yield 
the same set of six-fermion final states. In many studies of the linear 
collider prospects for top physics, single top quark production has been 
neglected. Notable exceptions are found in 
References~\cite{Boos:2001sj,Batra:2006iq}. 

A fully consistent analysis of the inclusive 
$e^+e^- \rightarrow W^{+} b W^{-} \bar{b}$ is currently impossible.
Even if event generation for the ILC TDR routinely included the full
$2 \rightarrow 6$ matrix element at leading order, higher-order 
corrections for this process are not available for the extraction
of top quark properties and couplings. In the continuum, results for
the cross-section of top quark pair production are available at NNLO.
Cross-sections for the process $W^+bW^-\bar{b}$ are for the moment only 
available at LO, but are required at least at NLO (note that NLO calculations 
are already available for the LHC).

In this note we investigate the impact of single top events in the study 
of top quark pair production and propose a more inclusive experimental
strategy that compares precise predictions for the 
$e^+e^- \rightarrow W^{+} b W^{-} \bar{b}$ process to measurements. 

\section{Distinguishing single top from top quark pair production}

As top quark pair production and single top quark production 
give rise to the same six-fermion final state, the question arises
as to how one can distinguish both sources. 
At a fundamental level the single top and top quark pair production 
processes are entangled by interference between the different diagrams. 
No algorithm can ever separate them fully. However, one could hope to use
some of the marked features of the $e^+e^- \rightarrow t\bar{t}$ process to
make it stand out among the other processes that give rise to the 
$W^+bW^-\bar{b}$ final state. One could then hope to isolate samples 
that are enriched in top quark pairs or single top quark events.

\begin{figure}[htb]
\begin{center}
  \includegraphics[width=0.8\linewidth]{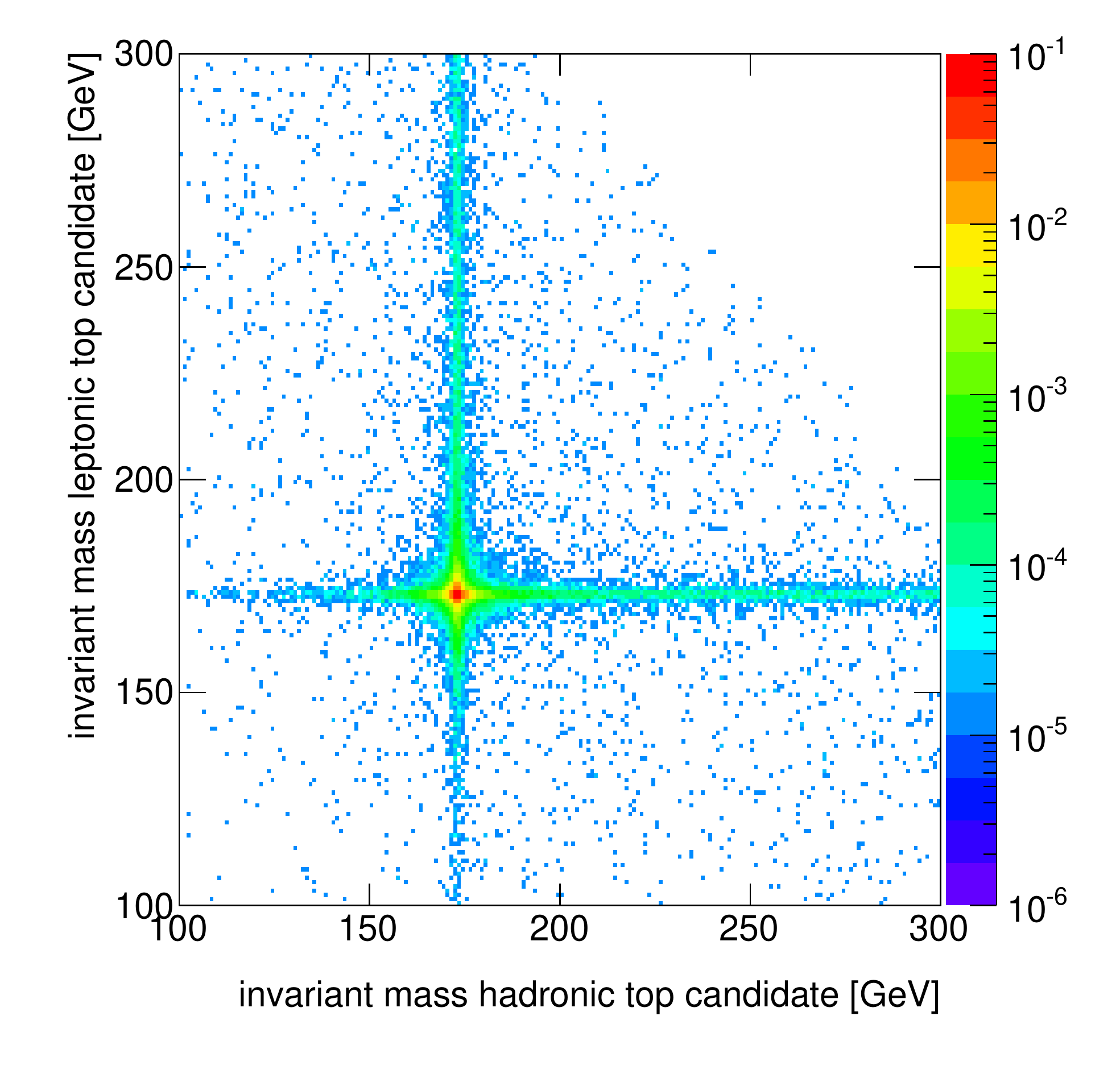}
  \caption{Reconstructed invariant mass at thruth level of the hadronic versus
           the leptonic side of $t\bar{t}\rightarrow W^+bW^-\bar{b}$ events. 
           The events cluster around the input top mass value, but one can
           observe a significant amount of off-shell events.
    \label{fig:invmass}}
\end{center} 
\end{figure}

In Figure~\ref{fig:invmass} we present the invariant 
mass of the $W^+b$ and $W^-\bar{b}$ combinations at truth level using
$e^+ e^- \rightarrow t\bar{t}\rightarrow W^{+} b W^{-} \bar{b}$ events 
generated with WHIZARD~\cite{Kilian:2014nya,Kilian:2007gr} at $\sqrt{s}=500$ 
GeV, including the effect
of initial state radiation and the beam energy spread expected at the ILC. 
The majority of events is found around $m_{W^+b} \sim m_{W^-\bar{b}} \sim m_t$. 
However, in a significant fraction of events one of the $W$-boson $b$-quark 
pairs has an invariant far from the top quark mass. The {\em cross} is
populated by a mixture of $t \bar{t}$ events with an off-shell top quark,
genuine single-top production through diagrams such as that in
the central panel of Figure~\ref{fig:diagrams}

\begin{figure}[htb]
\begin{center}
  \includegraphics[width=0.8\linewidth]{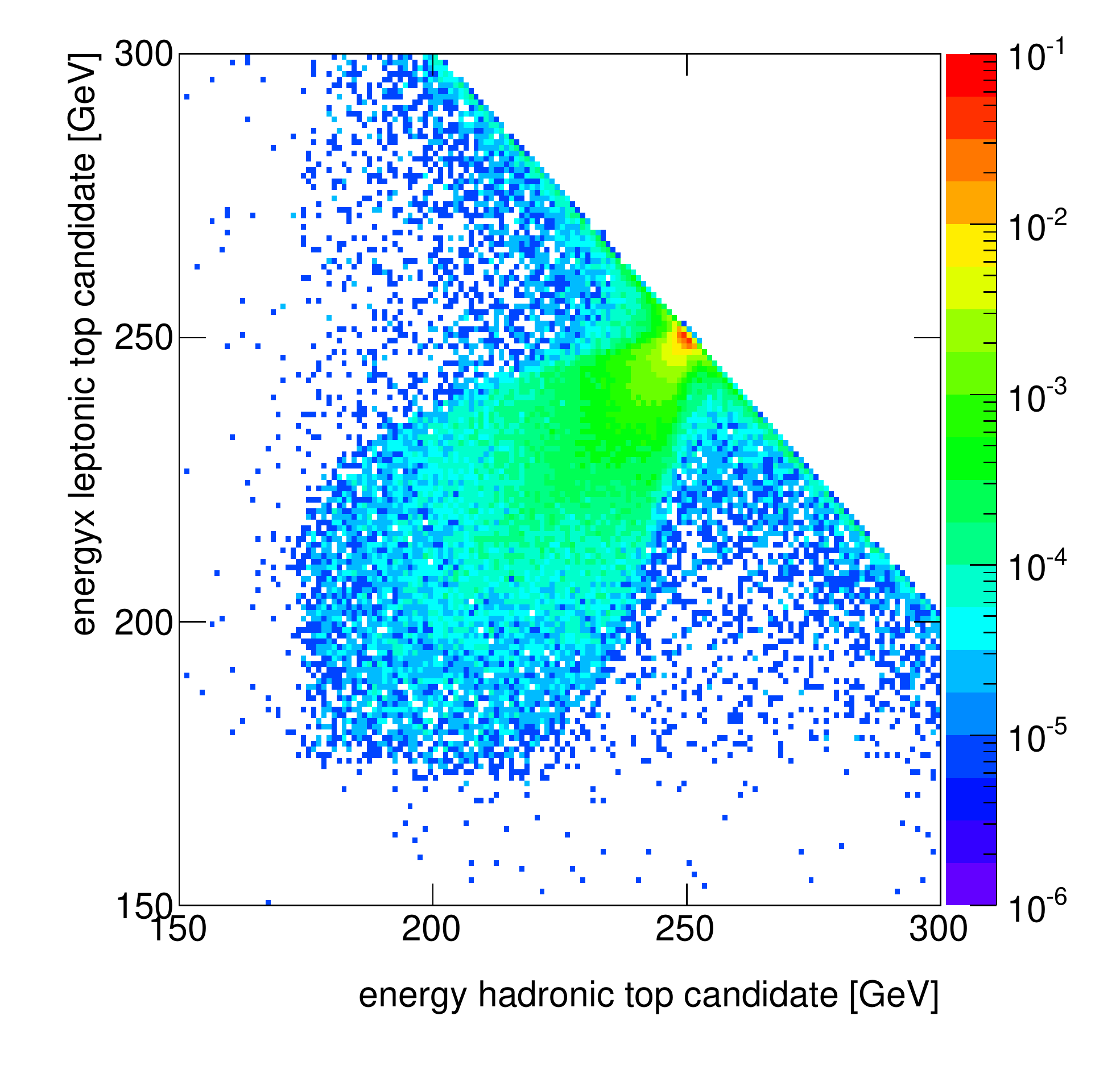}
  \caption{Reconstructed beam energy at thruth level of the hadronic versus the
           leptonic side of $W^+bW¯\bar{b}$ events. The center-of-mass energy is 500 GeV, so
           double-top events cluster at beam energies of 250 GeV. In single-top events,
           the beam energy follows the diagonal $E_{lep}+E_{had}=500$ GeV.           
    \label{fig:ebeam}}
\end{center} 
\end{figure}

In Figure~\ref{fig:ebeam} we present the reconstructed beam energy 
of the $Wb$ decaying leptonically versus the $Wb$ decaying hadronically.
The energy of the $Wb$ pair tends to cluster at about 
half the center-of-mass energy, as expected for double-top events.  
We observe however a significant fraction of events along the diagonal
$E_{lep}+E_{had}=\sqrt{s}$. Those are mostly single  single top events. 
This figure suggests another potential criterium for partial separation
of single and double-top events using the reconstructed beam energy.

These figures suggests an (ad-hoc) truth-level categorisation of events 
according to the number of on-shell top quark candidates, that is
used to quantify the non-$t \bar{t}$ contribution to $e^+ e^- \rightarrow t\bar{t}\rightarrow W^{+} b W^{-} \bar{b}$ production in the remainder of this paper. 
We consider the event to correspond to top quark pair production whenever the following relation is satisfied for both $Wb$ combinations:
\begin{equation}
|m_{Wb} - m^{MC}_t| < 15~\gev 
\label{eq:topcounting}
\end{equation}
Events that meet this criterium only for one of the $Wb$ pairs 
are labeled as single top quark events, and the remaining events are 
considered as non-top events.  

With this criterion we find that at 500 GeV typically 90\% of $e^+ e^- \rightarrow W^{+} b W^{-} \bar{b}$ events is formed by top quark pairs, 9\%
corresponds to single top production and only 1\% stems from processes
that do not involve top quarks. These fractions depend strongly on the
center-of-mass energy of the collider (more in Section~\ref{sec:energy}) and
on the beam polarization (more in Section~\ref{sec:500gev}). 
Over a broad range of energies and operating scenarios the fraction of 
single-tops is far from negligible and may have a significant impact on 
the measurement of top quark properties and the search for signs of new 
physics in $t\bar{t}$ production at lepton colliders.

\section{Experimental study at $\sqrt{s}= 500$ \gev{}}
\label{sec:500gev}

In this section we concentrate on the impact of realistic analysis cuts
on the composition of the 
$e^+ e^- \rightarrow W^{+} b W^{-} \bar{b}$ sample 
at $\sqrt{s} = $ 500 GeV. The analysis is based on the study of 
cross-sections and asymmetries at $\sqrt{s}= 500$ \gev{} in
Reference~\cite{Amjad:2013tlv}.
Final states of the type $l\nu q_1q_2b_1b_2$ are generated using the
WHIZARD Monte Carlo program. These events are mostly $W^+bW^-\bar{b}$ events, 
where one $W$ decays hadronically and the other leptonically. 
Some 100\,000 MC events are processed in a detailed simulation of
the ILD detector~\cite{ild} response based on GEANT4~\cite{geant4}.
Events are reconstructed, to the level of particle flow objects
and jets, using the ILC software package.
In these samples the beams are 100\% polarized. 

First, we discuss briefly the single top content as a funtion of beam 
polarization. The fraction of $W^+bW^-\bar{b}$ events that corresponds 
to single top quark production is quite sensitive to the polarization 
of the electron and positron beams. We evaluate the fractions of top quark 
pair, single top, and non-top events using the criterion of 
Equation~\ref{eq:topcounting} on $e^+ e^- \rightarrow el\nu q_1q_2b_1b_2$ 
samples generated with WHIZARD, including
initial state radiation and a realistic ILC luminosity profile.
For a fully left-handed 
electron beam and fully right-handed positron beam ($e_L^-e_R^+$) the total
cross-section is nearly three times larger. The top quark
pair fraction remains 90.2\%. The single top and non-top fractions
grow to 8.9\% and 0.9\%, respectively. For the opposite ($e_R^-e_L^+$) 
configuration, the total cross-section is similar to the unpolarized result.
The top quark pairs make up 94.3\% of the sample, with only 5.6\%
of single top quarks and about 0.1\% for non-top production.

The selection in Reference~\cite{Amjad:2013tlv} rejects background and
improves the fraction of well-reconstructed events using a cut on a 
$\chi^2$ formed by the reconstructed top mass, beam energy
and b--quark energy in the top rest frame. Only the hadronically decaying
top is used in the analysis. Table~\ref{table01} summarizes
the selection efficiencies at various stages of the analysis.

\begin{table}[htb] 
\caption{Selection efficiencies for various events samples. The efficiency estimates are based on 140.000 events for $e_L^-e_R^+$ and 40.000 $e_R^-e_L^+$ event and thus have a statistical error of several per mil.}
\begin{center} 

\begin{tabular}{|c|c|c|c|c|c|c|} 
\hline 
 & \multicolumn{3}{l}{$e_L^-e_R^+$} & \multicolumn{3}{l}{$e_R^-e_L^+$} \\
                    all   &  $t\overline t$      & single top    &  all   &  $t\overline t$      & single top       \\ 
\hline
Preselection                             &   47\% & 48\% & 44\%   & 49\% & 49\% & 47\% \\  
$\chi^2$ on hadronic top quark candidate & 31\%   & 33\%   & 19.0\% &  40\% & 40\% & 31\% \\ 
$\chi^2$ on both top quark candidates   & 32\%   & 34\%       & 21\% & 29\% & 29\% \\ 

\hline
\end{tabular}
\end{center} 
\label{table01} 
\end{table} 

\noindent
The preselection includes the lepton reconstruction and identification, 
the requirement of two b-tagged jets, and  .
The results in the second row are obtained after the cut on the 
hadronic top $\chi^2 < $ 15 employed in Reference~\cite{Amjad:2013tlv}. 
We observe that for both beam polarizations the efficiency for single top 
events is significantly smaller than for top quark pairs.
This is expected, since the $Wb$ pair mass and the beam energy are 
required to be compatible with the values expected for top quark pair production
in the  $\chi^2$ cut. However, the selection is insufficient to reduce
the single top {\em contamination} to a safe level. The fraction of single top
events continues to be large (5\% in the present example).
At this level it is expected to have a non-negligible effect on the 
cross-section measurement. An attempt to describe the remaining 
contamination using Monte Carlo events is likely to bring sizeable 
systematic uncertainties. The forward-backward
asymmetry is even more sensitive, since the asymmetry of single
top production is very small compared to that of top quark pair
production.

The inclusion of cuts on the leptonic top in the $\chi^2$ does not improve
the rejection. The third row in Table~\ref{table01} corresponds 
to cuts on both the hadronic and leptonic top $\chi^2$, where the $\chi^2$ cut 
is adjusted to obtain the same efficiency on the total sample.
 This can be understood as follows: once an event passes the
beam energy constraint, momentum conservation (forced by the neutrino
reconstruction) imposes a similar invariant mass on both the
hadronic and leptonic sides of the event, whether the event is
a single or a double top. In other words, the leptonic side
does not provide any additional background rejection.

\section{Analysis of top mass at threshold}

As pointed out a long time ago,  the cross-section of the
$e^+e^-\rightarrow t\overline t$ at threshold could be used to measure in a
precise and well defined way the top mass~\cite{Kuhn:1980gw}. Detailed
calculations are present in Reference~\cite{Fujii:1993mk} and 
a complete study of the extraction of top quark properties 
was performed in Reference~\cite{Martinez:2002st}. 

Recent reanalyses include realistic
beam energy spectra for the ILC and CLIC~\cite{Seidel:2013sqa} and
beam polarization~\cite{Horiguchi:2013wra}. In the
following we use the former analysis as a reference. The selection of events 
includes a kinematic fit, but as we showed in the previous section, the
leptonic side of the event brings no additional rejection of single top
events, since the neutrino can always be adjusted to fake a second top.
This analysis is very detailed, but single top events have not
been included. This paper includes for the case of ILC the program
TOPPIK, a NNLO calculation of the cross-section including the 1S
resonance, initial state radiation and the ILC luminosity spectrum
without beam polarization. For a mass input value of 174 \gev{}
and luminosity of 10 fb$^{-1}$ per data point, a fit is performed
in the range 344--354 \gev{} of center of mass energies. The result
is an impressive 27 MeV statistical error for the top mass,
and systematic errors well below 100 MeV (a recent paper~\cite{Marquard:2015qpa} determines 
the 4-loop correction for the conversion from $1S$ to the $\bar{MS}$ scheme; 
the related theory systematic could be as small as 10~\mev, thus prompting
a thorough evaluation of the experimental systematic uncertainties).

\begin{figure}[htb]
\begin{center}
  \includegraphics[width=0.9\linewidth]{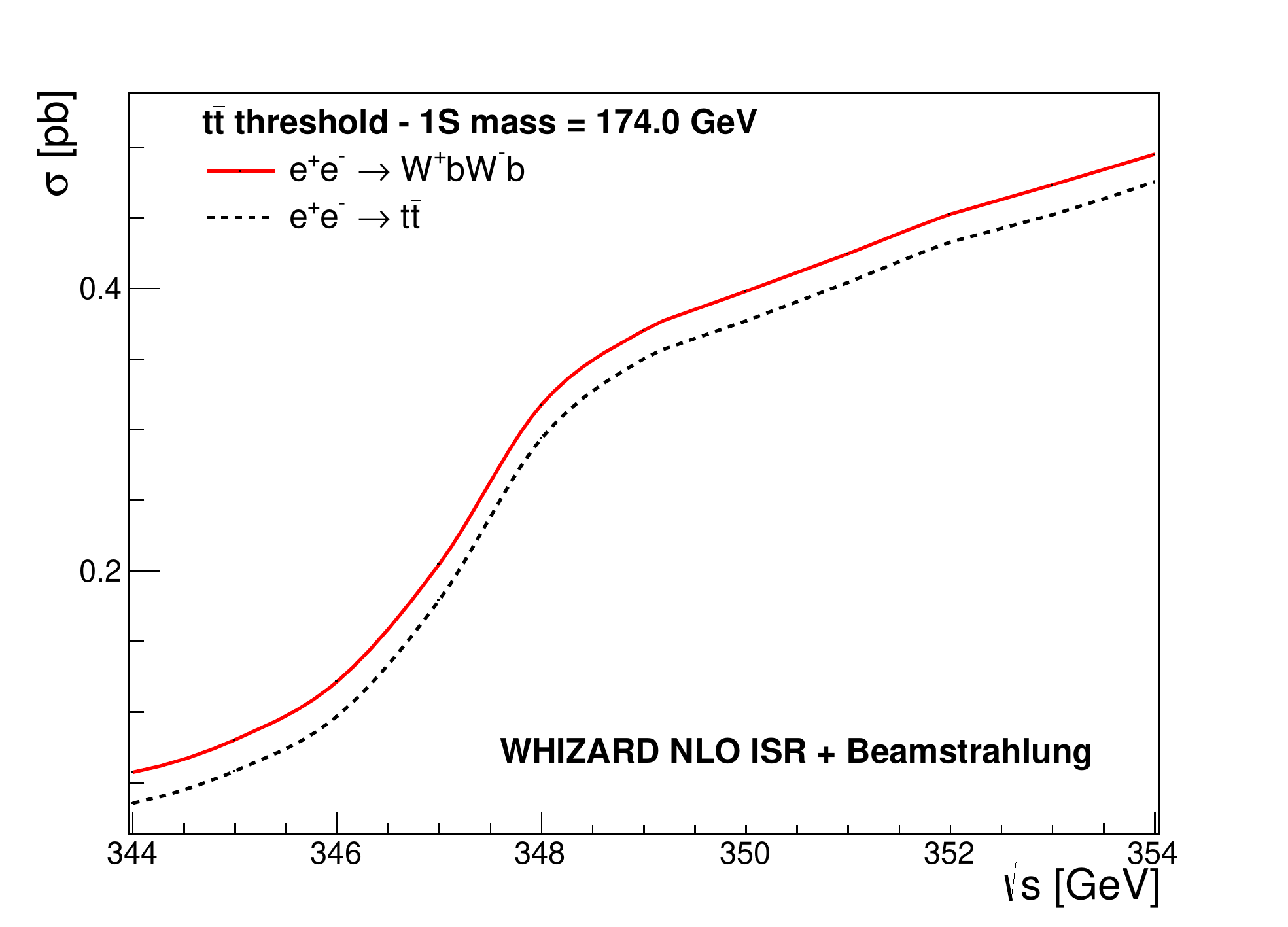}
  \caption{Cross-sections at NLO, according to WHIZARD, for $W^+bW^-\bar{b}$ and $t\bar{t}$ 
           production in $e^+e^-$ annihilations with unpolarized beams, as
           a function of center-of-mass energy, around the double-top
           threshold region. The 1S-top mass is set to 174 GeV.
    \label{fig:threshold}}
\end{center} 
\end{figure}

NLO calculations for the $W^+bW^-\bar{b}$ process have recently been
implemented in WHIZARD around the double-top production 
threshold including QCD bound state effects\footnote{J.Reuter and F.Bach, private communication.}.
We use these to estimate the single top fraction in $WbWb$ production. In 
Table~\ref{table02} and Figure~\ref{fig:threshold} the fraction of single 
top events is given for several
center-of-mass energies around the pair production thresholds.
The results include the effect of initial state radiation, of a realistic 
beam energy spectrum for the ILC at this energy, and the effect of polarized 
beams. The energy range and top quark mass follow the choices in 
Reference~\cite{Seidel:2013sqa}.

\begin{table}[htb] 
\caption{Single top content of the $W^+bW^-\bar{b}$ sample, for unpolarized
   and polarized beams, as a function of the center-of-mass energy.}
\begin{center} 
\begin{tabular}{|c|c|c|c|c|c|c|} 
\hline 
   $\sqrt{s}$  &   $e^-e+$     &    $e_L^-e_R^+$  &   $e^-e+$   &   $e^-e+$   &   $e^-e+$  & $e^-_Le+_R$  \\ 
    (GeV)      &    (LO)       &      (LO)       &    (NLO)   &    (NLO)  & (NLO, ISR, LS)  & (NLO, ISR, LS)   \\ 
\hline
   344      & 23\%   & 36\%    & 32\%    & 31.1\% & 38.2\% &  37.6\% \\ 
   345      & 19\%   & 30\%    & 22\%    & 22.3\% & 27.5\% &  27.8\%\\ 
   346      & 13\%   & 26\%    & 15\%    & 16.3\% & 20.1\% &  19.9\%\\ 
   347      &  9\%   & 19\%    &  9\%    & 9.9\% & 12.4\% &  11.5\%\\ 
   348      &  7\%   & 14\%    &  6\%    & 5.9\% & 7.5\% &  7.7\%\\ 
   349      &  5\%   & 10\%    &  5\%    & 4.4\% & 5.5\% &  6.1\%\\ 
\hline
\end{tabular}
\end{center} 
\label{table02} 
\end{table} 
 
\begin{figure}[htb]
\begin{center}
  \includegraphics[width=0.9\linewidth]{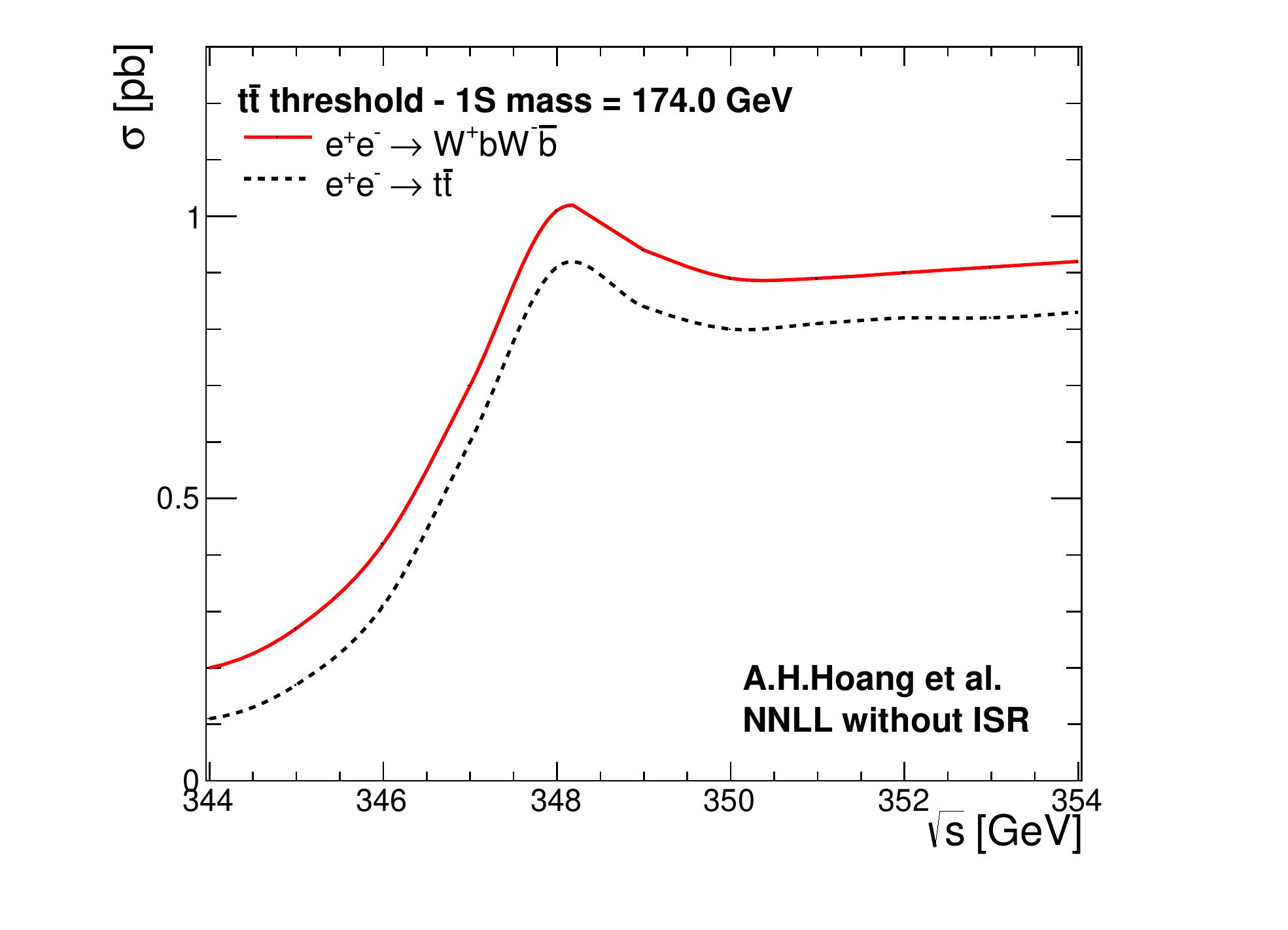}
  \caption{Cross-sections at NNLL, according to 
A.H.Hoang et al.~\cite{Hoang:2010gu}, for $t\bar{t}$ 
           production in $e^+e^-$ annihilations with unpolarized beams
           a function of center-of-mass energy. The 1S-top mass is set to 174 GeV.     \label{fig:threshold2}}
\end{center} 
\end{figure}

\begin{figure}[htb]
\begin{center}
  \includegraphics[width=0.9\linewidth]{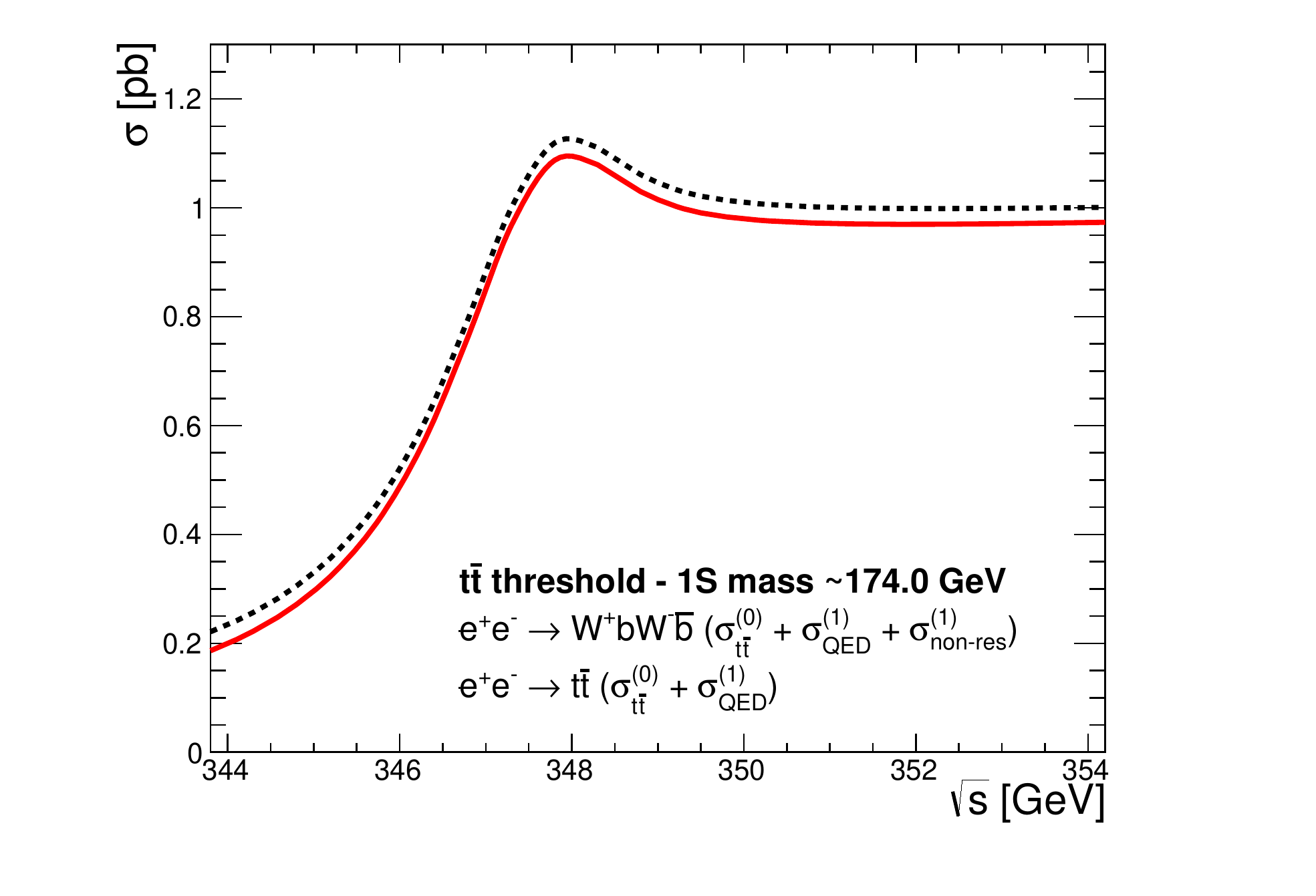}
  \caption{The effect on the cross-section of non-resonant corrections, 
according to M.~Beneke et al.~\cite{Beneke:2010mp}, for $e^+e^-$ annihilation to $W^+bW^-\bar{b}$ in the $t\bar{t}$ threshold region.
          The cross-section for unpolarized beams is given as a function of 
center-of-mass energy. With respect to the original publication, the curve is shifted such that the 1S peak coincides approximately with those of Figures~\ref{fig:threshold2} and~\ref{fig:threshold}. 
    \label{fig:threshold3}}
\end{center} 
\end{figure}

We note that the presence of single top events modifies 
significantly the cross-section. Moreover, the fraction
of single top events varies rapidly across the threshold region.
This does not mean, however,  that the top mass measurement
is affected. According to the method proposed in~\cite{Seidel:2013sqa},
a fit is performed to the cross-section measurement, using a predefined
function obtained from the double-top calculation at NNLO.
As long as the shape of the curve is not modified, the result is
not affected. We observe that according to the NLO calculation of 
the $W^+bW^-\bar{b}$ cross-section, the 
$t\bar{t}$ cross-section may be obtained quite exactly by shifting
down the $W^+bW^-\bar{b}$ cross-section by some 0.020 pb. 
There is only a residual shift at 348 GeV of 0.004 pb that would
shift the top mass by at most 20 MeV. The $W^+bW^-\bar{b}$ and
$t\bar{t}$ curves are displayed Figure~\ref{fig:threshold}.

This result is also supported by the 
NNLL result in Reference~\cite{Hoang:2010gu}, that we reproduce in
Figure~\ref{fig:threshold2}. The calculations are not directly
comparable, as initial state radiation and beam energy spread are not
included in Reference~\cite{Hoang:2010gu}. 
The authors find that the impact impact of a cut on the $Wb$ mass 
is an approximately constant shift. The cross-section for $t\bar{t}$ events 
with $Wb$ massin a range -15 $< m_{Wb} - m_t < $ 15 is, as before, shifted down 
from the inclusive  cross-section by 0.050 pb.
Finally, Figure~\ref{fig:threshold3} shows the effect of non-resonant
effects on the $WbWb$ cross-section. Again, the shift in the cross section
is essentially constant across the threshold region.

The single top content of the samples depends strongly
on the energy, as seen in table~\ref{table02}. In analyses where
the top quark is reconstructed and cuts are applied on the mass
and/or energy of the candidates, the selection
efficiency can be quite different for top quark pair and single
top events (up to 50\% smaller for single top in the example above). 
This means that, after correcting for efficiency,
the shape of the curve will be deformed if single top events are ignored.  
Note that this does not apply to Reference~\cite{Seidel:2013sqa}, since
in this analysis no selection cuts based on reconstructed
top quark mass and energy are applied.
 
As the most precise calculations are only available for the 
$e^+e^- \rightarrow t\bar{t}$
process, none of the studies of future 
collider prospects so far has taken into account the full
$W^+bW^-\bar{b}$  process. As soon as the necessary tools have been developed
we suggest to repeat the
study of the top quark mass extraction of future lepton
colliders using $W^+bW^-\bar{b}$ events and polarized beams.

\section{Energy dependence}
\label{sec:energy}

\begin{figure}[htb]
\begin{center}
  \includegraphics[width=1.0\linewidth]{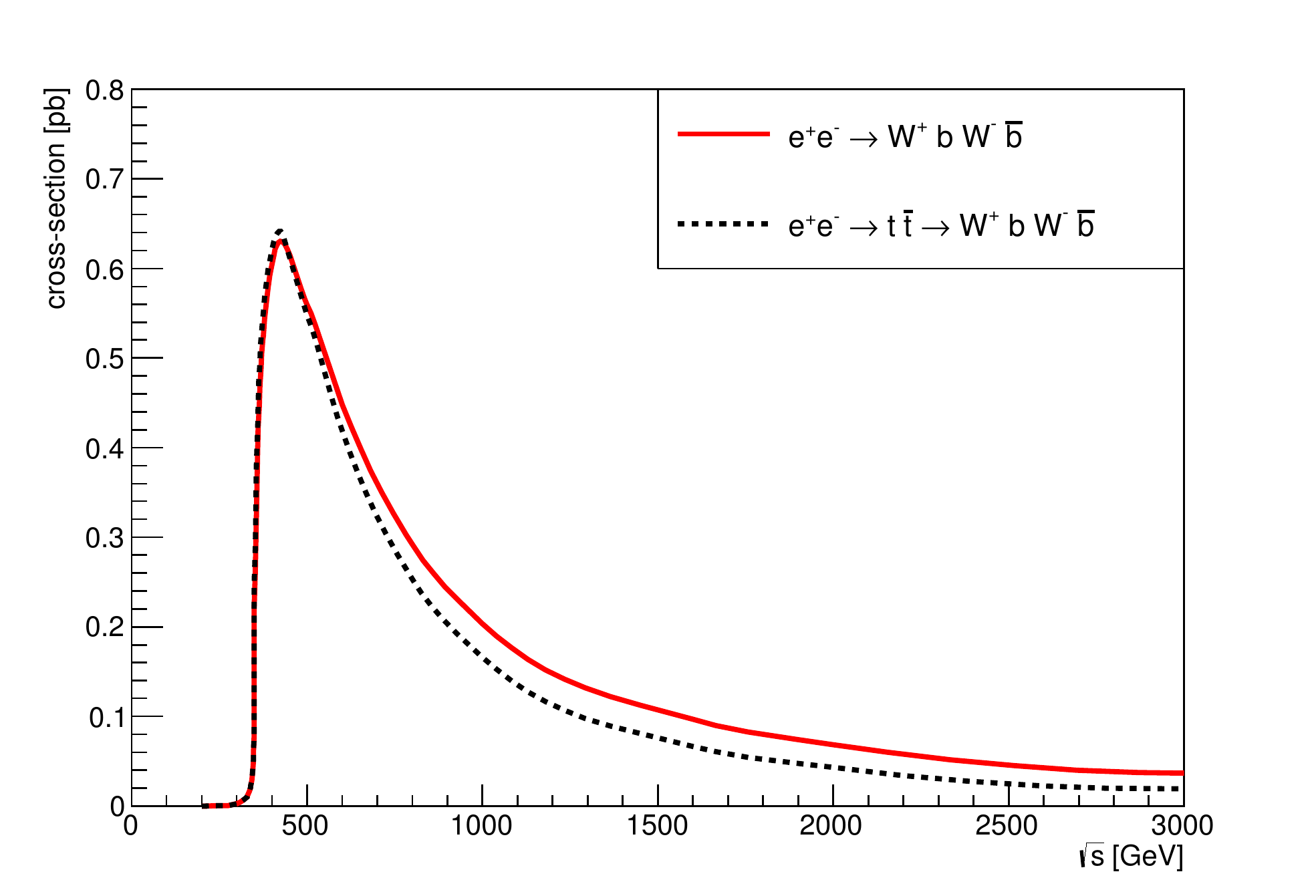}
  \caption{Cross-section for $W^+bW^-\bar{b}$ and $t\overline{t}$
           events produced in unpolarized electron-positron
           annihilations as a function of the 
           center-of-mass energy. The curves are obtained using
           WHIZARD at LO.
    \label{energydep}}
\end{center} 
\end{figure}

The composition of the $W^+bW^-\bar{b}$ sample is energy dependent
as shown in Fig.~\ref{energydep}, that presents the 
unpolarized LO cross-sections for $e^+e^- \rightarrow W^+bW^-\bar{b}$.
The solid line corresponds to the full 2 $\rightarrow$ 4 process, 
including single top production and non-top production. The dashed line
presents the results for the $e^+e^- \rightarrow t\bar{t} \rightarrow 
W^+bW^-\bar{b}$ process. All results have been obtained 
using WHIZARD. The difference between the $t\bar{t}$ 
cross-section and the full 2 $\rightarrow$ 4 result is 10\% at 500 \gev{} 
and rises to 20\% at 1~\tev{} and nearly 50\% at 3~\tev{}.

The relatively small contribution of the $e^+ e^- \rightarrow t\bar{t}$ process
at large center-of-mass energy is confirmed by an analysis of the number of 
on-shell top quarks ($|m_{Wb} - m^{MC}_t|<$ 15~\gev) in $W^+bW^-\bar{b}$ events 
at 3~\tev{}. The fraction of events with two on-shell top quarks  
is 48\%, events with a single on-shell top quark make up 35\% of the sample
and the remaining 17\% of events has no on-shell top quarks.
We conclude that while the rate for the $e^+e^-\rightarrow t\bar{t}$ process
drops at very large center-of-mass energy, single-top and non-top production
increase rapidly. The three processes become comparable in size for
$e^+e^-$ colliders operating in the multi-\tev{} regime.


\section{Summary and conclusions}

The prospects for precision top quark physics at future lepton colliders
relies on a comparison of measured cross-sections to very precise predictions 
of top quark pair production, including NLO and sometimes NNLO calculations. 
However, as we have shown in this note, single top events, that are 
ignored in most prospect studies, may lead to significant effects. 
A precise experimental separation of single top and top quark pair production
processes seems difficult. For this reason we advocate
the analysis of $W^+bW^-\bar{b}$ production, that includes single top
quark production, and leads to increased statistics and possibly smaller
systematic errors. To achieve the ultimate possible precision at the ILC,
observables for the $W^+bW^-\bar{b}$ final state must be calculated
with a precision well below 1\%, at NNLO, both at
production threshold and in the continuum.

\section*{Acknowledgement}

The authors thank the WHIZARD authors, in particular Juergen Reuter and Fabian Bach, for their continued support. We thank Pedro Ruiz-Femenia for helpful discussions.

\bibliographystyle{JHEP}
\bibliography{singletop}{}

\end{document}